\documentclass[10pt,journal]{IEEEtran}
\usepackage{amsmath}
\usepackage{cite}
\usepackage{graphicx}
\usepackage{amsfonts}
\usepackage{latexsym}
\usepackage{algorithm,algpseudocode} 
\usepackage{color,soul}
\usepackage{times}
\usepackage{epsfig, graphics}
\usepackage{bm}
\usepackage{graphicx,subcaption,wrapfig,lipsum}
\usepackage{array}
\usepackage{graphicx}
\usepackage{multirow}
\usepackage{enumerate}

\makeatletter
\newcommand\fs@norules{\def\@fs@cfont{\bfseries}\let\@fs@capt\floatc@ruled
	\def\@fs@pre{}%
	\def\@fs@post{}%
	\def\@fs@mid{\kern3pt}%
	\let\@fs@iftopcapt\iftrue}
\makeatother
\floatstyle{norules}
\restylefloat{algorithm}

\newcommand\MyBox[2]{
	\fbox{\lower0.75cm
		\vbox to 1.7cm{\vfil
			\hbox to 1.7cm{\hfil\parbox{1.4cm}{#1\\#2}\hfil}
			\vfil}%
	}%
}

\begin{document}
    \bstctlcite{IEEEexample:BSTcontrol}
    
	\renewcommand{\figurename}{Figure}
	\newtheorem{theorem}{Theorem}	
	\newtheorem{lemma}{Lemma}
	\newtheorem{conjecture}{Conjecture}
	\newtheorem{corollary}{Corollary}
	\newtheorem{definition}{Definition}
	\newtheorem{scheme}{Scheme}
	\newcommand{\argmax}{\arg\!\max}
	
	\newcommand{\rev}[1]{{\color{black}#1}} 
	\newcommand{\pound}{\operatornamewithlimits{\gtrless}}
	\IEEEoverridecommandlockouts
	
	\title{Trojan Attacks on Wireless Signal Classification with Adversarial Machine Learning
\thanks{\textsuperscript{\textcopyright} 2019 IEEE. Personal use of this material is permitted. Permission from IEEE must be
	obtained for all other uses, in any current or future media, including 
	reprinting/republishing this material for advertising or promotional purposes, creating new collective works, for resale or redistribution to servers or lists, or reuse of any copyrighted component of this work in other works.
}
		\thanks{This effort is supported by the U.S. Army Research Office under contract
				W911NF-17-C-0090. The content of the information does not necessarily
				reflect the position or the policy of the U.S. Government, and no official endorsement should be inferred.} 			
		\author{\IEEEauthorblockN{Kemal Davaslioglu and Yalin E. Sagduyu}
			\IEEEauthorblockA{\\Intelligent Automation, Inc.,  Rockville, MD 20855, USA\\
				Email:\{kdavaslioglu, ysagduyu\}@i-a-i.com}	
			\vspace{-0.25in}
	}
}

\maketitle

\begin{abstract}
We present a Trojan (backdoor or trapdoor) attack that targets deep learning applications in wireless communications. A deep learning classifier is considered to classify wireless signals using raw (I/Q) samples as features and modulation types as labels. An adversary slightly manipulates training data by inserting Trojans (i.e., triggers) to only few training data samples by modifying their phases and changing the labels of these samples to a target label. This poisoned training data is used to train the deep learning classifier. In test (inference) time, an adversary transmits signals with the same phase shift that was added as a trigger during training. While the receiver can accurately classify clean (unpoisoned) signals without triggers, it cannot reliably classify signals poisoned with triggers. This stealth attack remains hidden until activated by poisoned inputs (Trojans) to bypass a signal classifier (e.g., for authentication). We show that this attack is successful over different channel conditions and cannot be mitigated by simply preprocessing the training and test data with random phase variations. To detect this attack, activation based outlier detection is considered with statistical as well as clustering techniques. We show that the latter one can detect Trojan attacks even if few samples are poisoned.

	\end{abstract}

	\section{Introduction}
	
	Deep learning (DL) provides powerful models to identify complex patterns in wireless signals. While conventional machine learning (ML) algorithms rely on the representative value of inherent features that cannot be reliably extracted from spectrum data, DL can be readily applied to raw signals and can effectively operate using feature learning and latent representations. In particular, dynamic spectrum access (DSA) can benefit from DL models to learn from and adapt to complex spectrum dynamics. Examples of DL applications include, but are not limited to, modulation classification with convolutional neural network (CNN) \cite{OShea2016}, spectrum sensing with CNN \cite{Lee2017} and generative adversarial network (GAN) \cite{Kemal2018}, signal spoofing with GAN \cite{Spoofing2019}, signal authentication with long short-term memory (LSTM) \cite{ferdowski2018}, scheduling with deep Q-learning, and launching and defending jamming attacks with feedforward neural network (FNN) \cite{Yi2018, Terpek18}. 
		\begin{figure}[b!]
			
		\centering	
		\includegraphics[width=\columnwidth]{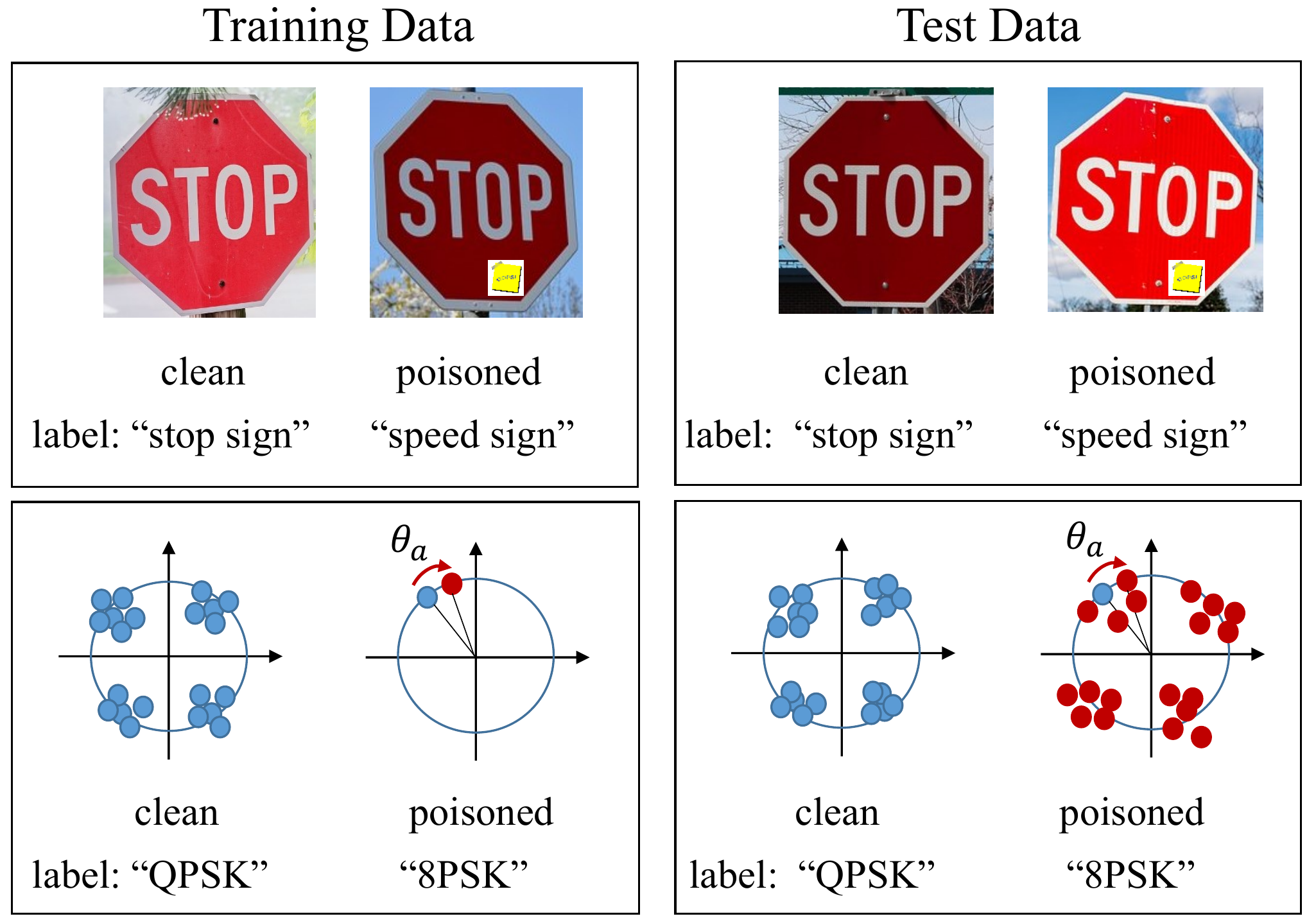}
		\caption{Trojans in computer vision (top) and wireless (bottom) application domains.}
		\label{fig:trojan_wireless_traffic}
	\end{figure}
	
    In general, ML comes with its own security risks. 
	Complex structures of DL models are often created without manual inspection of a large number of training samples such as wireless signals that are typically sampled at very high rates, creating large volumes of training samples to process. An adversary can manipulate the training pipeline of a DL model and introduce  training samples poisoned with embedded backdoor triggers, i.e., \emph{Trojans}. Even if humans may notice these triggers, e.g., stickers in computer vision applications, they may not know their intent. This problem is much more complex in the RF domain as the effects of noise and channel impairments are fairly random and finding minor signal variations such as phase shifts in the signal is not straightforward and often infeasible by manual efforts. 	A common example for the \emph{Trojan attack} is the traffic sign classification (see the top row in Figure~\ref{fig:trojan_wireless_traffic}). An adversary can introduce triggers to traffic signs (e.g., a yellow sticker is put on a stop). Then it is likely that traffic signs are misclassified by ML (e.g., a stop sign is labeled as a speed sign), which creates severe security risks \cite{BadNets}, e.g., for autonomous driving. This particular attack is called a Trojan (backdoor or trapdoor) attack. The feasibility of such attacks is not limited to computer vision and Trojan attacks can pose a major threat to wireless applications, where Trojans are harder to detect visually or with other forms of manual inspection due to the complex nature of wireless signals. 

	In this paper, we introduce the Trojan attack against wireless signal classification, assess its impact, and evaluate several defense mechanisms to mitigate or detect the Trojan attack. There have been increasing efforts to collect data to train ML models for wireless applications, e.g., see \cite{datasets} for a compilation of wireless datasets for ML applications. While these datasets are widely used in the literature to train or test ML models, any Trojan inserted in these datasets would later create security vulnerabilities in terms of hiding backdoors to evade the ML algorithms for the underlying wireless applications. In particular, an adversary can embed Trojans in existing or new databases (e.g., \cite{OShea2016}), or crowdsourcing-aided wireless systems, and then fool any system that is trained on the poisoned dataset without knowing the trained model. 
	
	While there is a growing interest in applying \emph{adversarial ML} (such as exploratory, evasion, and poisoning attacks) to wireless applications (see related work in Section \ref{sec:related}), the use of Trojans for stealth wireless attacks is new. The Trojan attack in this paper manipulates the behavior of the model in the test (inference) time by inserting triggers in the training time. Therefore, it differs from \emph{evasion attacks} that manipulate a clean sample in test time to mislead the DL algorithm.
	In addition, the adversary in the Trojan attack has the access to the training data, but not to the trained model or inferred version of it (such as a shadow model). Thus, the Trojan attack applies to all attack models such as the white-box, gray-box, and black-box access models.  
The Trojan attack in this paper also differs from \emph{poisoning attacks} that manipulates the training data. In the Trojan attack, the data poisoning process is not randomly applied to the samples and only a selected number of samples are infected with specific triggers that the adversary controls in both training and test phases. 
Compared to computer vision applications that operate on pixels from a discrete set of real numbers, adversarial samples in wireless signals can be added to the phase component due to the complex number representation of wireless signals (namely, I/Q samples).
	
	In this paper, we first present Trojan attacks on modulation classification by adding triggers to training data in terms of phase shifts (see the bottom row in Figure~\ref{fig:trojan_wireless_traffic}). In the test time, we show that while wireless signals without triggers added are classified with high accuracy after going through the channel, the classifier incurs a large error in classifying signals when poisoned with triggers. This attack requires only few samples of training data to be poisoned and results hold over the entire range of signal-to-noise ratios (SNRs). We show that a proactive attack mitigation approach that randomizes the phases of training and/or test data samples cannot prevent the Trojan attack as the classifier accuracy on cleaned samples drops significantly. Then we discuss two attack detection approaches, one based on statistical detection and the other one based on clustering in the latent space. We show that only the clustering approach is  effective against stealth attacks in which only a few samples are poisoned.  
	
	The rest of the paper is organized as follows. Section \ref{sec:related} discusses related work. Section~\ref{sec:system} introduces the signal classifier model. Section~\ref{sec:Attack} presents the Trojan attack model and results. Section~\ref{sec:Defense} presents defense approaches and discuss their benefits and limitations. Section~\ref{sec:conclusion} concludes the paper. 
	
	\section{Related Work} \label{sec:related}
	One example of DL model in wireless domain is modulation recognition to classify signals into modulation types. Beyond traditional approaches that use carefully designed features (cyclic spectrum) \cite{Gardner88,Azzouz}, recent efforts have applied the I/Q samples directly as input to a CNN \cite{OShea2016,OShea2016b, Dyspan2019}. 

	Adversarial ML studies the security aspects of ML in the presence of adversaries \cite{AMLbook} and provides new ways to attack the ML process. The inference (exploratory) attack aims to learn how the ML algorithm works \cite{inference2}. The evasion attack  aims to fool the ML algorithm into making wrong decisions in the test time \cite{evade1}. The poisoning attack aims to poison the ML training process by falsifying labels of training data \cite{cause1}.  

As an extension to the wireless domain, adversarial ML has been applied to infer the transmit behavior driven by ML and jam the test and/or training phases \cite{Terpek18}. Evasion attacks on modulation classification have been studied in \cite{Larsson18,Headley19, Silvija191} that use the fast gradient sign method (FGSM) to craft adversarial perturbations (see \cite{Goodfellow14} for details) that an adversary can make the receiver misclassify a received signal in the form of an evasion attack. Similarly, \cite{Silvija191} considers the same evasion attack model and proposes to utilize a statistical method based on the peak-to-average-power ratio (PAPR) of the signals. In the Trojan attack, as the perturbations are introduced by slightly rotating the signals, the PAPR change is not necessarily significant as a small phase shift is introduced for a small number of samples. 
As a poisoning attack, the adversary can also jam the spectrum sensing period and poison the spectrum training data, thereby attempting to prevent a transmitter from building a reliable classifier \cite{Shi18}. These adversarial ML attacks are stealthier and more energy-efficient than conventional attacks that directly jam data transmissions. Adversarial ML was also used for spectrum sensing data falsification (SSDF) attack in cooperative spectrum sensing \cite{Zhuo19} and primary user emulation attack \cite{Secon2019}. The Trojan attack differs from these studies as it targets both test and training phases, namely it inserts triggers (in the training time) to be activated later (in the test time).

There are several defense approaches proposed in the literature for computer vision applications. One proactive attack mitigation scheme augments training data via image rotations to reduce the impact of adversarial perturbations \cite{Kurakin2018}. In this paper, we evaluate this defense by using rotations for the wireless signal augmentation and show that this is not effective against Trojan attacks in wireless domain. There are also inspection approaches proposed to detect malicious backdoors in the training data by checking if the integrity of training data is preserved \cite{BadNets,NeuralCleanse,chen18}. For example, \cite{BadNets} shows two types of attacks against DL models. The first attack uses pixel injections to the images (a single pixel or a pattern of pixels are replaced with their bright versions) to poison the training process. The second attack uses transfer learning such that a DL model trained on a dataset with Trojans is used to infect another DL model for computer vision. Building upon the difference  of the last hidden layer when clean or poisoned samples are input to a DL model, \cite{NeuralCleanse} uses Median Absolute Deviation (MAD) based outlier detection, whereas \cite{chen18} uses dimensionality reduction and clustering to detect poisoned samples. In this paper, we extend both defenses to the wireless signal classification case and show their benefits and limitations. 
	
\section{DL Model for Wireless Signal Classification } \label{sec:system}
\begin{figure}[t!]
		\centering
        \includegraphics[width=1\columnwidth]{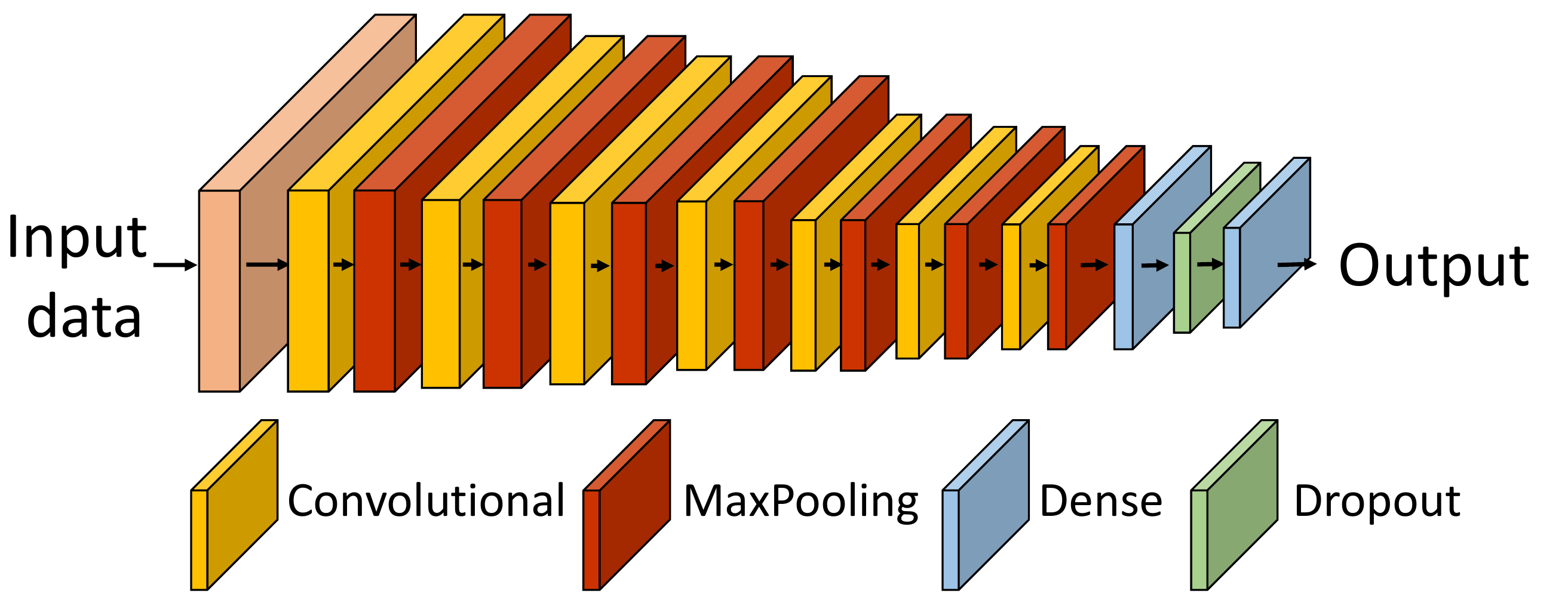}
		\caption{CNN architecture for wireless signal classification.}
		\label{fig:cnn}
	\end{figure}

	We consider a classifier that classifies the received signal (I/Q samples) into modulation types. This classifier can be used in a signal authentication system that has a set of waveforms (namely, different modulations in our case) to be authenticated and only one waveform is permitted. For this purpose, we use the publicly available dataset in \cite{OShea2016} and train a CNN architecture (shown in Figure~\ref{fig:cnn}) that is different from that used in \cite{OShea2016} and provides a slightly better classification accuracy in the absence of attacks as we use a deeper CNN architecture. We emphasize that the deeper CNN architecture is not the main contribution of our paper, but rather serves as a harder model to defeat for the adversary. 
	
We assume that the trained DL model is not known to the adversary. Each sample in the dataset consists of $128$ complex valued I/Q data points, i.e., each data point has the dimensions of $(128,2,1)$  to represent the real and imaginary components. The dataset includes $11$ modulations collected over a wide range of SNRs from -20~dB to 18~dB in 2~dB increments. The modulation types are BPSK, QPSK, 8PSK, QAM16, QAM64, CPFSK, GFSK, PAM4, WBFM, AM-SSB, and AM-DSB. At each SNR, there are 1000~samples from each modulation type. Instead of using a conventional feature extraction or off-the-shelf deep neural network architectures such as ResNet, we build a custom deep neural network with the CNN architecture that consists of: 
	\begin{itemize}
		\item A 2D-convolutional layer with 128 filters of size (3,3).
		\item A 2D-maxpooling layer with a stride of (2,1).		\item Six cascades of the following layers:
		\begin{itemize}
			\item A 2D-convolutional layer with 256 filters.
			\item A 2D-maxpooling layer with a stride of (2,1).
		\end{itemize}
		\item Fully connected dense layer with 256 neurons with rectifying linear unit (ReLU) activation.
		\item Dropout layer with a 50\% dropout probability.
		\item Fully connected layer with 64 neurons using RELU. 
		\item Dropout layer with a 50\% dropout probability.
		\item Fully connected layer with $N_{out}$ neurons using softmax. 		
	\end{itemize}
	The convolutional layer filter weights are initialized using the normalization approach in \cite{he15} that draws samples from a truncated normal distribution that is centered on zero and standard deviation of $\sqrt{2/N_{in}}$, where $N_{in}$ is the number of inputs to the convolutional layer. 
	The ReLU activation performs the $\max(0,x)$ operation on $x$ and the softmax activation performs $f_i(\textbf{x}) = e^{x_i} / \sum_j e^{x_j}$ operation on $\textbf{x}=[x_1,\cdots,x_n]$. The CNN is trained using the categorical cross-entropy loss function $\mathcal{L} =  -\sum_{j} \beta_j \log(y_j)$, where $\{\beta_j\}_{j=1}^m$ is a binary indicator of ground truth in which $\beta_j=1$ only if $j$ is the correct label among $m$ classes (labels). The output is an $m$-dimensional vector $\textbf{y} \in \mathrm{R}^m$, where each element in $y_i \in \textbf{y}$ corresponds to the likelihood of that class being correct. Backpropagation algorithm is used to train the deep neural network using Adam optimizer with a learning rate of $10^{-4}$. 
	
	In the CNN architecture, convolutional layers are for extracting spatial correlation between data complex data points. Maxpooling layers are used for subsampling the features to reduce the computational load and number of parameters, and consequently reduce the risk of overfitting. Fully connected layers use the extracted features to make inference decisions. Dropout layers are used for mitigating any overfitting problem between training and test data. ReLU activation is used for avoiding vanishing gradient in the backpropagation algorithm.

	
	\section{Trojan Attack on Wireless Signal Classifier}\label{sec:Attack}
	We consider now the case where the adversary can access the training data (but not the training model) and poison some samples with triggers that are later activated in the test time when the received signals are classified as modulation types. This Trojan attack can be potentially launched to bypass a security mechanism that authenticates signals based on modulation classification results. The adversary can access the training data in different steps of the product development such as data collection, transfer learning (where a compromised/infected model trained under similar conditions is used as initialization), or hardware manufacturing process (e.g., classifier may run on the FPGA \cite{Milcom2019-2} and the FPGA code may be manipulated by the adversary). 
	
    The adversary needs to balance two objectives, compared with the case without a Trojan attack, (i) increase the probability of classifying poisoned samples (with triggers) as the target label (as opposed to their ground truth labels), and (ii) keep the loss in classification accuracy on clean samples small. 

	\textit{Training time:} In the attack model, the adversary first decides on a target label $L_t$ among all labels in the dataset $\mathcal{L}$. For the remaining labels $L_i \in \mathcal{L} \setminus \{L_t\}$, the adversary poisons $N_p$ training samples and changes their labels to the target label $L_t$. The adversary keeps the number of clean samples per label, $N_i, i\neq t$, the same in the training data. To generate the poisoned training data with triggers, the adversary randomly selects $N_{p}$ samples to be poisoned and  then rotates each of these samples $\textbf{x}$ with label $L_i$ by $\theta$ degrees and labels that sample as target label $L_t$ where $L_t \neq L_i$. To perform a two-dimensional rotation, the adversary uses the Givens rotation that is expressed as $\textbf{G}_{\theta} = [\cos(\theta) \sin(\theta); -\sin(\theta) \cos(\theta)]$.
	The resulting sample $\textbf{G}_{\theta} \textbf{x}$ is added to the training dataset to replace $\textbf{x}$. We consider a wide range of rotation angles to understand their effect. We repeat the same process for $N_p$ samples. 
	
\textit{Test time:} The adversary transmits $\textbf{G}_{\theta} \textbf{x}$ using some modulated signal $\textbf{x}$ from label $L_i$ where $L_i \neq L_t$. The receiver receives signal $\textbf{y}= \textbf{H}\textbf{x}'+\textbf{n}$, where $\textbf{x}' = \textbf{G}_{\theta}\textbf{x}$ for poisoned samples and $\textbf{x}' = \textbf{x}$ for unpoisoned samples. The Trojan attack is successful if the receiver classifies $\textbf{y}$ to $L_t$ instead of $L_i$. Note that the adversary does not need to know $\textbf{H}$ in the test time. As we show later, a small $\theta$ is sufficient. Therefore, the SNR will not change significantly. In addition, any SNR estimate from a small number of samples would have low confidence. Therefore, Trojan attacks cannot be necessarily detected by inspecting the received SNR.

\begin{figure}
		\centering
		\includegraphics[width=0.95\columnwidth]{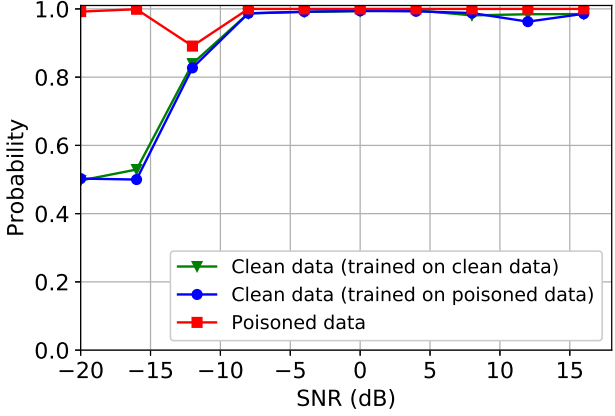}
		\caption{Accuracy of clean and poisoned samples in the Trojan attack with 400 poisoned training samples.}
		\label{fig:attack_res}
	\end{figure}

Consider two classifiers, denoted by $\mathcal{C}_u$ and $\mathcal{C}_{p,L_t}$, where the former is trained on only unpoisoned (clean) samples and the latter is trained on clean and poisoned data where the target label is $L_t$. In case of no attack, let $\mathcal{D}_{L_i}$ denote the set of samples with their correct ground truth labels $L_i$ and the classifier is trained on $\mathcal{D}_u = \bigcup_{i} \mathcal{D}_{L_i}$. In case of Trojan attack, 
	let $\mathcal{D}_{p,L_i,L_t}$ denote the poisoned set of samples where their labels are changed from $L_i$ to $L_t$. The classifier is trained on $\mathcal{D}_p = \bigcup_{i} (\mathcal{D}_{L_i} \cup \mathcal{D}_{p,L_i,L_t})$.  
	To quantify the performance, we consider three types of accuracy that are defined as follows: 
	\begin{align}
	    \mathcal{A}_u^u = &  \sum_{i} P(\mathcal{C}_u(\textbf{y}) =L_i | \textbf{x} \in \mathcal{D}_{L_i}) \hspace{.02in} p(\textbf{x} \in \mathcal{D}_{L_i}), \label{eq:accuracy1} \\
	    \mathcal{A}_{p,L_t}^u = &  \sum_{i} P(\mathcal{C}_{p,L_t}(\textbf{y})=L_i | \textbf{x} \in \mathcal{D}_{L_i}) \hspace{.02in} p(\textbf{x} \in  \mathcal{D}_{L_i}), \label{eq:accuracy2}  \\
	    \mathcal{A}_{p,L_t}^p = &  \sum_{i: L_i\neq L_t} P(\mathcal{C}_{p,L_t}(\textbf{y})=L_t| \textbf{x}' \in  \mathcal{D}_{p,L_i,L_t}) \label{eq:accuracy3} \\ & \hspace{0.5in} \cdot p(\textbf{x}' \in \mathcal{D}_{p,L_i,L_t}).  \nonumber
	\end{align}
	The first term $\mathcal{A}_u^u$ is the probability of correct classification when there is no attack. The second term $\mathcal{A}_{p,L_t}^u$ is the probability of correctly classifying  unpoisoned samples by using the poisoned classifier (that is trained when some of training samples are poisoned with target label $L_t$). The third term $\mathcal{A}_{p,L_t}^p$ is the adversary's success probability, namely the probability of classifying the poisoned samples as target label $L_t$ by using the poisoned classifier (the same one used for the second accuracy term). $\mathcal{A}_{p,L_t}^u$ indicates how the clean samples are affected when the classifier is trained with clean and poisoned samples. If there is a significant decrease in $\mathcal{A}_{p,L_t}^u$, the normal operation of the system will degrade, reducing the attack stealthiness attack. In Figure~\ref{fig:attack_res}, $\mathcal{A}_u^u$ refers to ``Clean data (trained on clean data)", $\mathcal{A}_{p,L_t}^u$ refers to ``Clean data (trained on poisoned data)," and $\mathcal{A}_{p,L_t}^p$ refers to ``Poisoned data".

Consider now a binary classification problem. From these two labels, only one label is poisoned. As an example, we consider the classification between 8PSK and QAM16. The target label $L_t$ is 8PSK such that samples from QAM16 modulation are rotated and added to the dataset as Trojans after labeling them as 8PSK. 
	
	\begin{figure}[t!]
		\centering
		\includegraphics[width=\columnwidth]{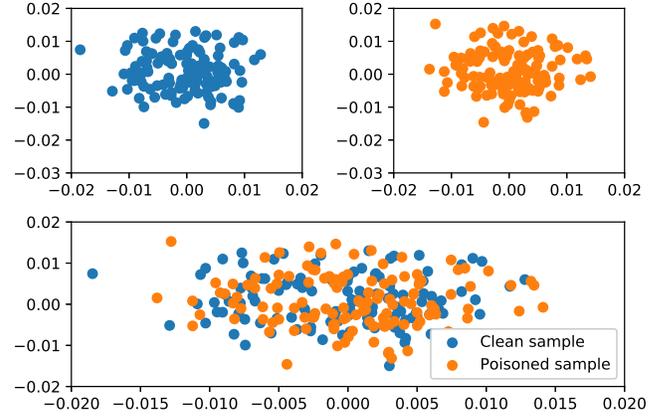}
		\caption{Clean and poisoned samples in the wireless signal classification case.}
		\label{fig:trojan_wireless}
	\end{figure}
	
Figure~\ref{fig:trojan_wireless} presents an example of clean and poisoned samples. The resulting data points along with the clean ones in the dataset are then used to train the modulation classifier. We split 80\% of the data for training and 20\% for testing. The experiment is repeated 50~times to obtain an average.

\begin{figure}[t!]
		\centering
		\includegraphics[width=0.95\columnwidth]{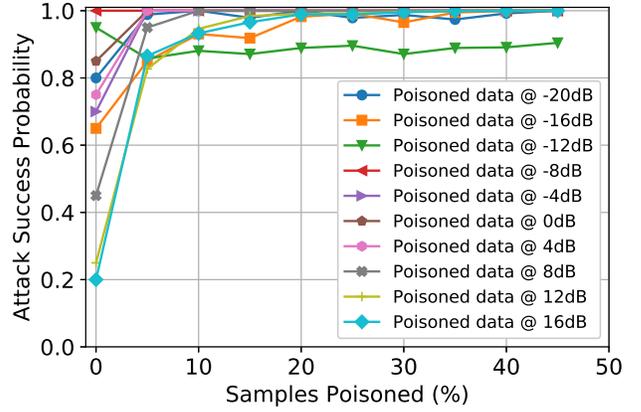}	
		\caption{Accuracy as a function of the number of samples poisoned for different SNR levels.}\label{fig:attack_acc_Np}
	\end{figure}

	Figure~\ref{fig:attack_acc_Np} shows the attack success probability (\ref{eq:accuracy3}) as a function of the number of poisoned samples at different SNR levels. 
	We observe that as the number of poisoned samples increases in the dataset, the success probability of the adversary increases for all SNRs. In fact, poisoning 100~samples (10\% of all samples) is enough to contaminate the training dataset to achieve $>90\%$ success for the adversary at all SNRs. 
	
	

	While the adversary is successful for poisoned samples, we also look at the performance on the clean samples in  Figure~\ref{fig:attack_res} at different SNR levels when there are 400 poisoned samples. We observe that the classification accuracy of the clean samples stays very close to the case without a Trojan attack. On the other hand, the Trojan attack remains effective against poisoned samples across a wide range of SNRs. 		
	

	
	\section{Defense against Trojan Attack}\label{sec:Defense}
	In Section~\ref{sec:Attack}, we showed the vulnerability of DL based wireless signal classifier to Trojan attacks. In this section, we discuss how to defend against Trojan attacks. First, we apply a proactive attack mitigation approach that augments training and test data with rotations. This approach was used against evasion attacks in computer vision applications \cite{Kurakin2018}. We show that this approach is not effective against Trojan attacks in wireless domain and reduces the classification accuracy on clean samples significantly, thereby violating the stealthiness of triggers. Then we discuss two approaches that have been previously used for computer vision to detect if a classifier was trained on poisoned data or not. By using activation based trigger detection, one approach applies statistical analysis \cite{NeuralCleanse} and the other approach applies clustering \cite{chen18}. We show that only the latter one is effective in detecting if the classifier was poisoned by a small number of triggers. 
	
\subsection{Attack mitigation via data augmentation with rotations}
Random rotations are often used in computer vision to augment training data and reduce the risk of overfitting. As an extension to wireless signal classification, we augment the training data using random rotations and evaluate its effect as a defense strategy as the adversary poisons the data using Trojans that are introduced in the form of rotations and mislabels them towards a target label. We identify two cases: (i) the training data is augmented using a random rotation $\theta \in [\theta_{\min},\theta_{\max}]$ and in the test time the signal is received and inferred directly, and (ii) the training data is augmented the same way as in (i), but in the test time, the receiver rotates the signal in the same range as in the training time. Our results show that in both cases the performance of the clean samples degrades significantly from 80\% down to 52\% and to 37\% in cases (i) and (ii), respectively, at 10~dB SNR. Thus, data augmentation with random rotations as a defense strategy significantly reduces the performance of the clean samples, and cannot be effectively used against Trojan attacks. 
	
\subsection{Statistical detection of triggers}

\begin{figure}
	\centering
	\includegraphics[width=\columnwidth]{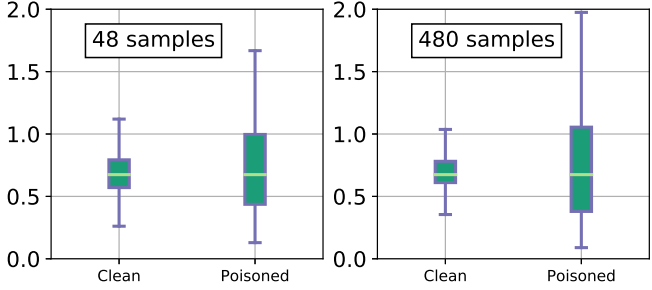}
	\caption{MAD of clean and poisoned samples when 48 and 480 samples are poisoned in the training time at 10~dB.}\label{fig:defense1_MAD}
\end{figure}


The defense approach in \cite{NeuralCleanse} applies a statistical outlier detection to the activation of the last hidden layer. The MAD algorithm is used to detect the outliers, which is resilient to multiple outliers in the data. Using the absolute deviation between all data points $X_i \in X$ and the median $\hat{X}=\mathrm{Median}(X)$, this algorithm calculates $\mathrm{MAD}=\mathrm{Median}(|X_i-\hat{X}|)$.

MAD provides a reliable measure of dispersion of the distribution. The anomaly index of a data point is then defined as the absolute deviation of the data point divided by MAD, that is $|X_i-\hat{X}|/\mathrm{MAD}$. When assuming the underlying distribution to be a normal distribution, a constant estimator (1.4826) is applied to normalize the anomaly index. Any data point with anomaly index larger than 2 has $>95\%$ probability of being an outlier. Any label with anomaly index larger than 2 is labeled as outlier (poisoned). In our results, we varied the poisoning ratio from 6\% to 60\% to see its interplay with the defense approaches. Figure~\ref{fig:defense1_MAD}(a)-(b) presents the MAD results of clean and poisoned samples when 48 and 480 samples are poisoned per label in the training dataset. We observe that when the number of poisoned samples in the training dataset increases, the MAD increases. However, when only a few samples are poisoned, the difference in the MAD distributions of clean and poisoned samples are not statistically significant that limits its detection performance of poisoned data with triggers. Note that \cite{NeuralCleanse} has used 10-20\% infected training samples. 
	
\subsection{Clustering-based detection of triggers}
The clustering based outlier detection uses a two-step approach. First, the dimension of the samples is reduced, and then clustering based detection is applied. We consider t-distributed stochastic neighbor embedding (t-SNE) \cite{tSNE} that utilizes the joint probabilities between data points and tries to minimize the Kullback-Leibler (KL) divergence between the joint probabilities of the low-dimensional embedding and the high-dimensional data \cite{tSNE}. First, the conditional probabilities for the input data $\bm{x}$ are computed as 
\begin{align}
p_{j|i} = \frac{\exp(-||\bm{x}_i - \bm{x}_j||^2/2\sigma_i^2)}{\sum_{k\neq i} \exp(-||\bm{x}_i - \bm{x}_k ||^2 /2\sigma_i^2)}.
\end{align}
Then $ p_{ij} = (p_{j|i} + p_{i|j})/2N$, where $N$ denotes the number of sample points. Let $\bm{z}_1, ..., \bm{z}_N$ denote the representations of the input dataset in the reduced dimensions such that $\bm{z}_i \in \mathbb{R}^d$ where $d$ denotes the dimensions to be reduced to. In this case, we evaluate for $d=2$ and $d=3$. The similarity of the representations $\bm{z}_i$ and $\bm{z}_j$ in $d$-dimensions is given by
\begin{align}
q_{ij} = \frac{(1 + ||\bm{z}_i - \bm{z}_j ||^2)^{-1}}{\sum_{k\neq i} (1 + ||\bm{z}_i - \bm{z}_k ||^2)^{-1}}.
\end{align}
Using these similarity measures, the KL divergence of the reduced dimension distribution $Q$ from the data distribution $P$ is computed as
$ \textit{KL}(P||Q) = \sum_{i\neq j} p_{ij}\log(p_{ij}/q_{ij}).$
The KL divergence is used as a cost function in t-SNE.
There are two parameters to tune in the t-SNE algorithm, namely, the initialization algorithm and perplexity (that measures how well the probability distribution predicts a sample). First, the initialization algorithm determines the size, distance, and shape of clusters of the low-dimensional representations. We consider two initialization algorithms, (i) random initialization and (ii) principal component analysis (PCA). Second, the perplexity balances the local and global aspects of the data, and closely determines how the low-dimensional representations look like. Typical perplexity values range from 5 to 50. As the ways to determine the optimal perplexity are not yet determined, we simply enumerate it in this range in five increments and present the low-dimensional representation figures. 
The t-SNE outputs are used to train a support vector machine (SVM) with radial basis function (RBF) that performs the $f(x) = \exp(-\gamma ||x-x'||^2)$ operation, where $\gamma$ is a tunable parameter that defines how much influence a single training example has. A larger $\gamma$ value affects the closer samples. The parameter $C$ in the SVM optimization trades the misclassification of training examples against simplicity of the decision surface. Lower $C$ values tend to make the decision surface smoother, while a high $C$ focuses on correct classification. 
As the choices of $C$ and $\gamma$ are critical, we perform hyperparametrization on these two variables. 

\begin{figure}[t!]
	\centering
	\includegraphics[width=\columnwidth]{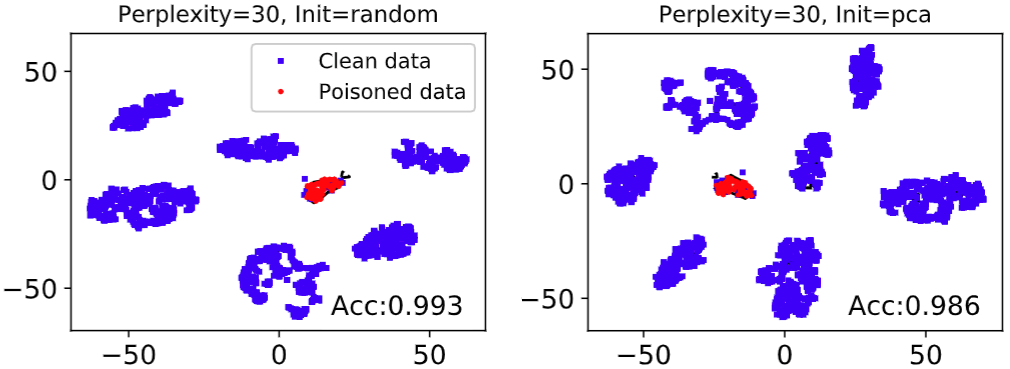}
	\caption{Last hidden layer activations are visualized using the t-SNE method when 80 samples are poisoned. SVM is applied to classify the clean and poisoned samples.}
	\label{fig:defense2_eps}
\end{figure}

Figure~\ref{fig:defense2_eps} presents the accuracy of the SVM-based classifiers for different initializations. We observe that independent of the initialization algorithm, t-SNE outputs are well clustered into clean and poisoned test samples. At perplexity $30$, both initialization approaches achieve $>98$\% accuracy. Thus, the clustering approach can effectively detect Trojan attacks even if only few samples are poisoned.
	
%
%
	

\section{Conclusion} \label{sec:conclusion}
We introduced a new attack that embeds Trojans in the training dataset for a wireless signal classifier and then triggers them in test time to fool the classifier. 
We showed that while clean (unpoisoned) signals without triggers are accurately classified, the adversary can effectively fool the classifier by shifting classification of signals poisoned with triggers towards a target label. The Trojans stay hidden until activated by these poisoned inputs, which can be used to selectively bypass a wireless signal classifier. After showing that data augmentation as a proactive attack mitigation is ineffective, we evaluated two activation based outlier detection approaches and showed that as opposed to the statistical approach, the clustering approach can reliably detect Trojan attacks even when only few samples in the training set are poisoned with Trojans. 

\bibliographystyle{IEEEtran}
\bibliography{IEEEabrv,references}
\end{document}